\documentclass[a4paper]{article}
\usepackage[affil-it]{authblk}
\usepackage{epsfig}\usepackage{latexsym}\usepackage{amssymb}\usepackage{bm}
\usepackage{float}
\usepackage{subfigure}
\usepackage{amsmath}
\usepackage{amsfonts}
\usepackage[usenames]{color} 
\newcommand{\be}{\begin{equation}} \newcommand{\ed}{\end{displaymath}}
\newcommand{\bd}{\begin{displaymath}} \newcommand{\ee}{\end{equation}}
\newcommand{\bea}{\begin{eqnarray}} \newcommand{\eea}{\end{eqnarray}}
\newcommand{\ba}{\begin{array}} \newcommand{\ea}{\end{array}}

\begin{document}

\title{A `Dysonization' Scheme for Identifying Quasi-Particles using Non-Hermitian Quantum Mechanics}

\author{Katherine Jones-Smith%
  \thanks{Electronic address: \texttt{kas59@physics.wustl.edu}; Corresponding author}}
\affil{Physics Department\\ Washington University in Saint Louis\\ 1 Brookings Drive\\ Saint Louis, MO 63130\\ USA}

\date{Received 2 April 2012; Accepted 15 May  2012}

\maketitle

\abstract{In 1956 Dyson analyzed the low-energy excitations of a ferromagnet using a Hamiltonian that was non-Hermitian with 
respect to the standard inner product. 
This allowed for a facile rendering of these excitations (known as spin waves) as weakly interacting bosonic quasi-particles. 
More than 50 years later,  we have the full denouement 
of non-Hermitian quantum mechanics formalism at our disposal when considering Dyson's work, both technically and contextually. Here we recast Dyson's work on ferromagnets explicitly in terms of two inner products, with respect to which the Hamiltonian is always self-adjoint, if not manifestly `Hermitian'. Then we extend his scheme to doped antiferromagnets described by the $t-J$ model, in hopes of shedding light on the physics of high-temperature superconductivity. }


\section{Introduction}\label{sec1}
A major goal in condensed matter physics is to represent the low-energy physics
of strongly interacting quantum many-body systems in terms of weakly interacting
quasiparticles that are either bosonic or fermionic \cite{bib1}. In a seminal
paper Dyson \cite{bib5} showed that a Heisenberg ferromagnet could be represented
as a theory of weakly interacting bosons called magnons or spin waves; this representation 
allowed thermodynamic calculations of unprecedented accuracy.

Dyson's formulation had the unorthodox feature that the bosons were governed by a Hamiltonian
that was superficially non-Hermitian. More precisely there were two inner products at work
in Dyson's representation of a ferromagnet. First, there was what we will call the ``kinematic
inner product'' with regard to which  the boson creation and annihilation operators were 
adjoints of each other. In other words, this was the inner product with regard to which the
quasiparticles were bosons. Second there was the ``dynamical inner product'' with regard
to which the Hamiltonian was self-adjoint. Conversely, however, the quasiparticles were 
not bosonic with respect to the dynamical inner product and the Hamiltonian was not 
self-adjoint with respect to the kinematic inner product. 

By contrast the conventional approach is far more restrictive in that there is only a single
inner product with regard to which the quasiparticles are defined and with regard to which
the Hamiltonian and all other physical operators must be self-adjoint. In this paper we 
explore whether Dyson's more flexible concept of non-Hermitian quasiparticles can be more
broadly applied, particularly to problems that have so far resisted conventional Hermitian 
analysis. 

The $t-J$ model is believed to capture the essential physics of the cuprate super-conductors,
which represent one of the grand unsolved puzzles of theoretical physics \cite{bib9}. In this chapter we 
apply non-Hermitian quantum mechanics to this model and obtain a representation of its
low energy physics in terms of a Dyson boson and a Dyson fermion. By design these
quasiparticles are defined with respect to a kinematic inner product; the Hamiltonian that governs
them is not self-adjoint with respect to the kinematic inner product but with respect to the
dynamical inner product. 
An outline of the paper is as follows. 
First we review Dyson's work on ferromagnets,
highlighting the role of the two inner products. We then adapt the analysis to antiferromagnets,
a useful prelude to the study of the $t-J$ model. In the following section we describe
a spin $s$ generalization of the $t-J$ Hamiltonian (the physical case relevant to the cuprates
is $s= 1/2$). A natural and convenient way to write the $t-J$ Hamiltonian is to use a super-algebra
that is a super-symmetric generalization of the su(2) angular momentum algebra \cite{bib11}. After presenting
this supersymmetric formulation of the $t-J$ model we finally write the problem in terms of 
non-Hermitian quantum mechanics. The presentation here closely follows that in \cite{bib7}. 
\section{Magnets}
%

\subsection{Single spin}

A single spin has $2s+1$ basic states $|s, m>$ where $s$ is
the total spin and $m$ is its $z$-component. $s$ is the same for all states of the multiplet
and $m= -s, \ldots, s$. These states are assumed to be orthonormal
\begin{equation}
< s, m | s, m' > = \delta_{m m'}.
\label{eq:spinorthonormality}
\end{equation}
The spin-operators $S_z, S_+$ and $S_-$ obey the angular momentum algebra
\begin{equation} 
[ S_+, S_-] = 2 S_z, \hspace{2mm} 
[ S_+, S_z] = - S_+, \hspace{2mm}
[ S_-, S_z] = S_-,
\label{eq:angularmoment}
\end{equation}
where, as usual, the spin-raising operator $S_+ = S_x + i S_y$ and the spin-lowering operator
$S_- = S_x - i S_y$. As shown in textbooks, the effect of these operators on the basis states
$|s, m>$ is 
\begin{eqnarray}
S_+ |s, m> & = & (s - m)^{1/2} (s + m + 1)^{1/2} |s, m+1>, \nonumber \\
S_- |s, m> & = & (s - m + 1)^{1/2} (s + m)^{1/2} |s, m-1>, \nonumber \\
S_z |s, m> & = & m |s, m>.
\label{eq:raisingspin}
\end{eqnarray}

Dyson introduced an alternative set of basis states
\begin{equation}
|u> = F_u |s, - s + u>
\label{eq:dysonspinbasis}
\end{equation}
where $u =0, \ldots, 2s$. The state $|0>$ corresponds to having the $z$-component of the 
spin maximally down; the states $|1>, |2>, |3>, \ldots$ correspond to raising the $z$-component
by increments of one. These states are orthogonal but not normalized
\begin{equation}
< u | v > = F_u^2 \delta_{u,v}.
\label{eq:dysonfu}
\end{equation}
The normalization factors $F_0 = 1$ and 
\begin{equation}
F_{u} = \left( 1 \left[ 1 - \frac{1}{2s} \right] \left[ 1 - \frac{2}{2s} \right]
\ldots \left[ 1 - \frac{u-1}{2s} \right] \right)^{1/2}
\label{eq:dysonfu}
\end{equation}
for $u = 1, 2, \ldots 2s$. $F_u$ is judiciously chosen to map the spin-raising operator 
$S_+$ to the bose creation operator $b^{\dagger}$, as will be seen below. 

Making use of eqs (\ref{eq:raisingspin}), (\ref{eq:dysonspinbasis}) and (\ref{eq:dysonfu}) it is not
difficult to show 
\begin{eqnarray}
S_+ |u > & = & \sqrt{2s} \sqrt{u+1} |u + 1>, \nonumber \\
S_- |u > & = & \sqrt{2s} \left[ 1 - \frac{u-1}{2s} \right] \sqrt{u} | u - 1>, \nonumber \\
S_z |u> & = & (-s + u) |u>.
\label{eq:dysonspinraise}
\end{eqnarray}

Now consider a different Hilbert space with two operators $b$ and $b^{\dagger}$ that are
the adjoints of each other under a certain inner product, the ``kinematic inner product''. These operators
are assumed to satisfy the bose commutation relations
\begin{equation}
[ b, b^{\dagger}] = 1.
\label{eq:bose}
\end{equation}
Provided the kinematic inner product is positive definite it follows inexorably by standard textbook
arguments that the basic states in this Hilbert space form an infinite ladder $|u)$ with $u = 0, 1, 2, \ldots$
The state $|0)$ has the defining characteristic
\begin{equation}
b |0) = 0;
\label{eq:bosevacuum}
\end{equation}
we say this is a state with zero bosons. The state
\begin{equation}
|u) = \frac{1}{\sqrt{u!}} (b^{\dagger})^u |0)
\label{eq:stateudefnd}
\end{equation}
is said to contain $u$ bosons. These states are 
orthonormal under the kinematic inner product 
\begin{equation}
( u | v )_{{\rm kin}} = \delta_{u,v}
\label{eq:kinematicorthonormality}
\end{equation}
and the effect of the bose creation and annihilation operators on these states is
\begin{eqnarray}
b^{\dagger} |u) & = & \sqrt{u+1} |u+1), \nonumber \\
b |u) & = & \sqrt{u} |u-1), \nonumber \\
b^{\dagger} b |u) & = & u |u).
\label{eq:bosetrinity}
\end{eqnarray}

Following Dyson, we now establish a mapping between the space of spins and the bose
oscillator space by identifying the spin state $|u>$ with the boson state $|u)$. Thus
\begin{equation}
|u> \rightarrow |u)
\label{eq:dysonmapping}
\end{equation}
for $u = 0, \ldots, 2s$. States with more than $2s$ bosons have no spin space counterpart.

Dyson's mapping allows us to export the inner product of the spin space to the bose space.
We call this induced inner product the dynamical inner product. Explicitly
\begin{equation}
(u | v)_{{\rm dyn}} = F_u^2 \delta_{u,v}
\label{eq:dysonamical}
\end{equation}
for $u = 0, \ldots, 2s$. We take $F_u = 0$ for $u>2s$. Thus states with more than $2s$ bosons
are ``weightless''.

Dyson's mapping eq (\ref{eq:dysonmapping}) also allows us to establish the following
correspondence between spin and bose operators
\begin{eqnarray}
S_+ & \rightarrow & \sqrt{2s} b^{\dagger}, \nonumber \\
S_- & \rightarrow & \sqrt{2s} \left[ 1 - \frac{ b^{\dagger} b}{2 s} \right] b, \nonumber \\
S_z & \rightarrow & - s + b^{\dagger} b.
\label{eq:firstdysonrepresentation}
\end{eqnarray}
This correspondence follows from comparison of eq (\ref{eq:dysonspinraise}) and (\ref{eq:bosetrinity}).
$b$ and $b^{\dagger}$ are not the adjoints of each other under the dynamical inner
product. Since we are denoting the adjoint with respect to the kinematic inner product as
$^{\dagger}$, let us signify the adjoint with respect to the dynamical inner product by $^{\star}$.
We can then see for example that 
\begin{equation}
(b^{\dagger})^{\star} = \left[ 1 - \frac{b^{\dagger} b}{2s} \right] b
\label{eq:starrelation}
\end{equation}
and 
\begin{equation}
(b^{\dagger} b)^{\star} = b^{\dagger} b.
\label{eq:staralso}
\end{equation}
\subsection{Heisenberg Ferromagnet}

We now consider a two-dimensional Heisenberg ferromagnet in which the spins
occupy the sites of a square lattice. Thus the lattice sites $(m,n)$ have position vector
${\mathbf r}_{mn} = m a \hat{{\mathbf e}}_x + n a \hat{{\mathbf e}}_y$
where $\hat{{\mathbf e}}_x$ and $\hat{{\mathbf e}}_y$ are unit vectors along the
$x$ and $y$ axes, $m$ and $n$ are integers, and $a$ is the lattice constant. Each
site has four nearest neighbors. The site $(m,n)$ has neighbors located at 
${\mathbf r}_{mn} + {\boldsymbol \delta}$ where ${\boldsymbol \delta} = a \hat{{\mathbf e}}_x$, 
$a \hat{{\mathbf e}}_y$, $- a \hat{{\mathbf e}}_x$ and $- a \hat{{\mathbf e}}_y$ respectively for
the four neighbors. We denote the spin operator at position ${\mathbf r}$ as $S_+({\mathbf r}), S_-({\mathbf r})$
and $S_z ({\mathbf r})$. Operators at a given site are assumed to obey the angular momentum 
algebra eq (\ref{eq:angularmoment}); spin-operators at different sites are assumed to commute.
We consider a spin $s$ ferromagnet so the basic states at each site are a spin multiplet of $2 s + 1$ states.
The Hamiltonian for a Heisenberg ferromagnet is
\begin{eqnarray}
H_F & = & - \frac{J}{2} \sum_{{\mathbf r}} \sum_{{\boldsymbol \delta}} 
[ S_z ({\mathbf r}) S_z ({\mathbf r} + {\boldsymbol \delta} ) + 
\nonumber \\
& & \frac{1}{2} S_+({\mathbf r}) S_-({\mathbf r} + {\boldsymbol \delta} ) +
\frac{1}{2} S_+({\mathbf r} + {\boldsymbol \delta})  S_-({\mathbf r}) ]. \nonumber \\
\label{eq:heisenbergferro}
\end{eqnarray}
Thus each spin is coupled to its nearest neighbors. We assume the exchange constant $J > 0$. 

Now consider a system of bosons $b({\mathbf r})$ and $b^{\dagger} ({\mathbf r})$ that live on a
square lattice in two dimensions (lattice constant $= a$). The operators $b({\mathbf r})$ and $b^{\dagger}({\mathbf r})$
are assumed to be adjoints of each other under the kinematic inner product. They are assumed to obey the
bosonic commutation relation
\begin{equation}
[ b({\mathbf r}), b^{\dagger} ({\mathbf r'}) ] = \delta_{{\mathbf r}, {\mathbf r'}}.
\label{eq:latticebosons}
\end{equation}
Thus $b^{\dagger} ({\mathbf r})$ creates bosons at site ${\mathbf r}$; 
$b({\mathbf r})$ annihilates them. We may now represent the ferromagnetic Heisenberg Hamiltonian
eq (\ref{eq:heisenbergferro})
in terms of bosonic quasiparticles by using Dyson's mapping. From the correspondence
eq (\ref{eq:firstdysonrepresentation}) between spin and bose operators 
we obtain the bosonic form of the Heisenberg 
Hamiltonian
\begin{eqnarray}
{\cal H}_F & = & \frac{Js}{2} \sum_{{\mathbf r}, {\boldsymbol \delta}} 
[ 2 b^{\dagger} ({\mathbf r}) b({\mathbf r}) - b^{\dagger} ({\mathbf r}) b ({\mathbf r} + {\boldsymbol \delta})
- b^{\dagger} ( {\mathbf r} + {\boldsymbol \delta} ) b ({\mathbf r}) ]
\nonumber\\
& + & \frac{J}{4} \sum_{{\mathbf r}, {\boldsymbol \delta} } 
[b^{\dagger} ({\mathbf r}) b^{\dagger} ({\mathbf r} + {\boldsymbol
\delta}) b^2 ({\mathbf r} + {\boldsymbol \delta}) + 
b^{\dagger} ({\mathbf r}) b^{\dagger} ({\mathbf r} + {\boldsymbol \delta}) b^2 ({\mathbf r}) ]
\nonumber \\
& - & \frac{J}{2} \sum_{{\mathbf r}, {\boldsymbol \delta}} b^{\dagger} ({\mathbf r}) b ({\mathbf r}) 
b^{\dagger} ({\mathbf r} + {\boldsymbol \delta} ) b({\mathbf r} + {\boldsymbol \delta} ).
\label{eq:dysonferro}
\end{eqnarray}

Note that the boson Hamiltonian ${\cal H}_F$ is not self-adjoint under the kinematic inner product
(${\cal H}_F^{\dagger} \neq {\cal H}_F$) due to the terms in the second line of eq (\ref{eq:dysonferro}). 
However it is self-adjoint under the dynamical inner product
(${\cal H}_F^{\star} = {\cal H}_F$).

\subsection{Heisenberg Anti-ferromagnet}

A Heisenberg anti-ferromagnet is simply a ferromagnet with $J < 0$. An equivalent but more
convenient description of the Heisenberg anti-ferromagnet on a square lattice is the following: Imagine two
interpenetrating square lattices, the site labelled $(m,n)$ on the first lattice is located
at ${\mathbf r}_1 (m,n) = m a \hat{{\mathbf e}}_x + n a \hat{{\mathbf e}}_y$. Here $m$ and $n$
are integers. The sites of the second
square lattice are displaced from those of the first by $(a/2) \hat{{\mathbf e}}_x + (a/2) \hat{{\mathbf e}}_y$.
Thus the site labelled $(m,n)$ on the second lattice is located at ${\mathbf r}_2 = (m + 1/2) a \hat{{\mathbf e}}_x +
(n + 1/2) a \hat{{\mathbf e}}_y$. Regardless of the lattice on which it sits, each site has four nearest
neighbors. The displacements from a given site to its four nearest neighbor sites are 
${\boldsymbol \delta}_1 = (a/2) \hat{{\mathbf e}}_x + (a/2) \hat{{\mathbf e}}_y$, 
${\boldsymbol \delta}_2 = (a/2) \hat{{\mathbf e}}_x - (a/2) \hat{{\mathbf e}}_y$, 
${\boldsymbol \delta}_3 = - (a/2) \hat{{\mathbf e}}_x + (a/2) \hat{{\mathbf e}}_y$, and
${\boldsymbol \delta}_4 = - (a/2) \hat{{\mathbf e}}_x - (a/2) \hat{{\mathbf e}}_y$.
We imagine there is a spin at each site and that the spin at each site is antiferromagnetically
coupled to its nearest neighbors. Thus the Hamiltonian for a Heisenberg anti-ferromagnet  is
\begin{eqnarray}
H_A & = & J \sum_{{\mathbf r}, {\boldsymbol \delta}} [ S_z^{(1)} ({\mathbf r}) S_z^{(2)} ({\mathbf r} + {\boldsymbol 
\delta}) + 
\nonumber \\
& & 
\frac{1}{2} S_+^{(1)} ({\mathbf r}) S_-^{(2)} ({\mathbf r} + {\boldsymbol \delta}) + 
\frac{1}{2} S_+^{(2)} ({\mathbf r} + {\boldsymbol \delta}) S_-^{(1)} ({\mathbf r})]. 
\nonumber \\
\label{eq:heisenbergantiferro}
\end{eqnarray}
The sum over ${\mathbf r}$ in eq (\ref{eq:heisenbergantiferro}) extends over the sites of the
first lattice; the sum over ${\boldsymbol \delta}$ extends over the four nearest neighbor
displacements enumerated above. The superscripts $^{(1)}$ and $^{(2)}$ over the spin
operators serve to remind us that the spin is on lattice one or lattice two respectively. 	


For the Heisenberg ferromagnet the exact ground state is that all the spins point maximally
down along the $z$-axis\footnote{Or along any other direction. The ground state of a ferromagnet
spontaneously breaks rotational symmetry and thus there is a manifold of equivalent ground
states.}. In Dyson's boson representation the ferromagnetic ground state is the state in which no 
bosons are present. Anti-ferromagnets present an altogether more formidable problem. 
The exact ground state for an anti-ferromagnet is not known except in one dimension for the
case of spin $s = 1/2$. The ideal `N\'{e}el state' is one in which the spins on the first lattice are
maximally down along the $z$-axis and the spins on the second lattice are maximally up along the
$z$-axis. The N\'{e}el state is not the exact ground state of the anti-ferromagnet but it is 
believed to be qualitatively similar\footnote{There are many circumstances where it is known
the N\'{e}el state is not even qualitatively right: in one dimension, on a triangular lattice in
two dimensions or even on a square lattice in two dimensions if next nearest neighbor
interactions act to frustrate N\'{e}el ordering.} and therefore a good starting point from 
which to obtain a more accurate picture of the ground and excited states of a Heisenberg
anti-ferromagnet. Thus in representing an anti-ferromagnet in terms of Dyson bosons we 
shall take the N\'{e}el state to be the one with no bosons present. 

To this end we establish a second mapping between a single spin and a single bose
oscillator. In this second ``anti-Dyson'' mapping a state with spin maximally up is to be 
identified with the state of zero bosons. Thus we introduce the anti-Dyson basis for a spin 
multiplet
\begin{equation}
| u, A > = G_u |s, s - u > 
\label{eq:antidysonbasis}
\end{equation}
where $u = 0, \ldots, 2s$. The normalization constant $G_0 = 1$ and 
\begin{equation}
G_u = \left( 1 
\left[ 1 - \frac{1}{2s} \right] 
\left[ 1 - \frac{2}{2s} \right] \ldots
\left[ 1 - \frac{u-1}{2s} \right] \right)^{-1/2}
\label{eq:gu}
\end{equation}
where $u = 1, 2, \ldots, 2s$. $G_u$ has been judiciously chosen 
to ensure that the spin raising operator $S_+$ maps to the bose annihilation 
operator $b$, as will be seen below. 

Making use of eq (\ref{eq:raisingspin}), eq (\ref{eq:antidysonbasis}) and 
(\ref{eq:gu}) it is not difficult to show 
\begin{eqnarray}
S_+ | u; A > & = & \sqrt{2s} \sqrt{u} | u - 1; A >, 
\nonumber \\
S_- | u; A > & = & \sqrt{ 2 s } \sqrt{u+1} \left[ 1 - \frac{u}{2s} \right] | u + 1; A >,
\nonumber \\
S_z | u; A > & = & (s - u) | u; A >.
\label{eq:antidysonspin}
\end{eqnarray}
We may  now establish an anti-Dyson mapping between spins and bose oscillators
by identifying the spin state $|u; A>$ with the bose oscillator state $|u)$. Thus
\begin{equation}
| u; A > \rightarrow |u)
\label{eq:antidyson}
\end{equation}
for $u = 0, \ldots, 2s$. States with more than $2s$ bosons have no spin space counterpart. 
The anti-Dyson mapping allows us to export a dynamical inner product to the bose space
as before. The remarks made earlier about this dynamical inner product apply mutatis mutandis
[see the paragraph surrounding eq (\ref{eq:dysonamical})].

The anti-Dyson mapping eq (\ref{eq:antidyson}) also allows us to establish a second 
correspondence between spin and bose operators
\begin{eqnarray}
S_+ & \rightarrow & \sqrt{2s} b,
\nonumber \\
S_- & \rightarrow & \sqrt{2s} b^{\dagger} \left[ 1 - \frac{b^{\dagger} b}{2 s} \right],
\nonumber \\
S_z & \rightarrow & s - b^{\dagger} b.
\label{eq:seconddyson}
\end{eqnarray}
This correspondence follows from comparison of eqs (\ref{eq:antidysonspin}) and (\ref{eq:bosetrinity}).

Equipped with the second Dyson mapping we now return to the Heisenberg anti-ferromagnet. We consider
two interpenetrating square lattices as above and assume that there are two kinds of lattice bosons. One kind
lives on the sites of the first lattice: $b_1^{\dagger}({\mathbf r}_1)$ creates this kind of boson at site
${\mathbf r}_1$; $b_1 ({\mathbf r}_1)$ annihilates it. The other kind live on the second lattice and 
are created and annihilated by $b_2^{\dagger} ({\mathbf r}_2)$ and $b_2({\mathbf r}_2)$ respectively.
These creation and annihilation operators are adjoints of each other with respect to the kinematical
inner product and are assumed to obey bosonic commutation relations
\begin{equation}
[ b_i({\mathbf r}), b_j ({\mathbf r'}) ] = \delta_{{\mathbf r}, {\mathbf r}'} \delta_{ij}
\label{eq:twobosons}
\end{equation}
where $i$ and $j$ equal $1$ or $2$. 

We may  now represent the Hamiltonian for the Heisenberg Hamiltonian eq (\ref{eq:heisenbergantiferro})
in terms of bosonic quasi-particles using Dyson's mapping between spins and bosons, eq 
(\ref{eq:firstdysonrepresentation}) on the sites of the first lattice and using the anti-Dyson mapping
eq (\ref{eq:seconddyson}) on the sites of the second lattice. This strategy ensures that the N\'{e}el
 state
corresponds to the boson vacuum and yields a bosonic form of the Heisenberg Hamiltonian
\bea
{\cal H}_A&  = &J s \sum_{{\mathbf r}, {\boldsymbol \delta}} [ b_1^{\dagger} ({\mathbf r}) b_1 ({\mathbf r}) +
b_2^{\dagger} ({\mathbf r} + {\boldsymbol \delta}) b_2 ({\mathbf r} + {\boldsymbol \delta}) ]
\nonumber \\
&+ &
J s \sum_{{\mathbf r}, {\boldsymbol \delta}} [ 
b_1^{\dagger} ({\mathbf r}) b_2^{\dagger} ({\mathbf r} + {\boldsymbol \delta}) +
b_2 ({\mathbf r} + {\boldsymbol \delta}) b_1 ({\mathbf r}) ]
\nonumber \\
&- &  
J \sum_{{\mathbf r}, {\boldsymbol \delta}} 
b_1^{\dagger} ({\mathbf r}) b_1 ({\mathbf r}) b_2^{\dagger} ({\mathbf r} + {\boldsymbol \delta}) 
b_2 ({\mathbf r} + {\boldsymbol \delta})
\nonumber \\
&- &  \frac{J}{2} \sum_{ {\mathbf r}, {\boldsymbol \delta}} 
b_1^{\dagger} ({\mathbf r}) b_2^{\dagger} ({\mathbf r} + {\boldsymbol \delta})
b_2^{\dagger} ({\mathbf r} + {\boldsymbol \delta}) b_2 ({\mathbf r} + {\boldsymbol \delta}) 
\nonumber \\
&- &  \frac{J}{2} \sum_{ {\mathbf r}, {\boldsymbol \delta}} 
b_1^{\dagger} ({\mathbf r}) b_1 ({\mathbf r}) b_1 ({\mathbf r}) b_2 ({\mathbf r} + {\boldsymbol \delta}).
\label{eq:antidysonboson}
\eea
Note that the boson Hamiltonian ${\cal H}_A$ is not self-adjoint under the kinematic inner product
(${\cal H}_A^{\dagger} \neq {\cal H}_A$) due to the terms in the last two lines of eq (\ref{eq:antidysonboson}).
However it is self-adjoint under the dynamical inner product (${\cal H}_A^{\star} = {\cal H}_A$). A Hamiltonian 
of this form was introduced and analyzed in ref \cite{bib6}.
\section{Doped Magnets}

A typical cuprate such as La$_{2-x}$Sr$_x$CuO$_4$ consists of stacked planes of Cu atoms.
Within a plane the Cu atoms are arranged in a square lattice. In the pure compound La$_2$CuO$_4$
there is one electron available per Cu atom. If electron-electron interactions were weak the electrons
could hop from atom to atom via tunneling. However in the cuprates the electron-electron repulsion is
strong, forbidding double occupancy of the Cu sites. Each site is therefore occupied by a single electron.
The electrons are locked in place and immobile. A material like this is called a `Mott insulator'. The only
degree of freedom is the electron spin that can point up or down at each site. The decidedly unequal
competition between hopping and electron-electron repulsion tends to make the spins align  
anti-ferromagnetically. The undoped cuprates may therefore described by the antiferromagnetic
Heisenberg Hamiltonian. (See for example \cite{bib10}.) 

In the doped compound La$_{2-x}$Sr$_x$CuO$_4$ there are only $1-x$ electrons per site and therefore
a fraction $x$ of the sites are unoccupied. The absence of electrons (``holons'') can hop and when the 
density of holons is sufficiently high the materials are observed to exhibit strange metallic and then 
superconducting behavior. The competition between hopping and electron-electron repulsion for
the doped compounds is described by the $t-J$ Hamiltonian. In the next section the $t-J$ Hamiltonian
is formulated in a way that is particularly well suited to our present purpose. 

\subsection{Supersymmetric formulation of $t-J$ Model}

In the parent compound there are two possible states for each site: spin up or spin down. 
In the doped material each site has three possible states: spin up, spin down or missing electron.
The missing electron state corresponds to zero spin and a positive charge $+e$ on the site. In the following
it will be useful to consider a spin-$s$ generalization wherein there are $4s+1$ states per site. 
The site may either be in one of the $2s+1$ states $|s, m>$ with $m=-s, \ldots, s$ or in one of the
$2s$ states $|s-1/2,m>$ with $m = - (s-1/2), \ldots, s-1/2$. If the site is in a spin $s$ state,
$|s, m>$, the total spin is $s$, the $z$-component of the spin is $m$ and the site is 
assumed to have no charge. On the other hand if it is in a spin $s-1/2$ state, $|s - 1/2, m>$,
the total spin is $s-1/2$, its $z$-component is $m$ and the site has a positive charge $+e$ due
to the lack of one electron. In summary, whereas the basic states per site of a spin $s$ magnet are a single 
spin $s$ multiplet $|s,m>$, the basic states per site for our $t-J$ model are a ``super-multiplet'': a pair
of multiplets with spin $s$ and spin $s-1/2$. The physically relevant case is $s = 1/2$. 

Having specified the basic states at each site we must now describe the basic operators out
of which the $t-J$ Hamiltonian will be built. For a magnet these operators are 
$S_+, S_-$ and $S_z$. They satisfy the su(2) angular momentum algebra eq (\ref{eq:angularmoment})
and their action on the states $|s, m>$ of a spin $s$ multiplet is well-known eq (\ref{eq:raisingspin}).
Now it turns out there is a super-algebra that is a natural generalization of the su(2) algebra and
the $t-J$ model can be written (super)naturally in terms of the elements of this algebra; this appears to have 
been first
noted by Weigmann \cite{bib11},  and subsequently solved exactly in one dimension by Bares and Blatter \cite{bib2a}.

The super-algebra has eight elements. Six of them are raising and lowering operators (also known
as Weyl elements): $S_+, S_-, R_+,
R_-, T_+$ and $T_-$. The remaining two are the Cartan elements $A$ and $S_z$. Since this is a
super-algebra the elements may also be grouped differently into commuting elements 
($S_+, S_-, S_z, A$) and anti-commuting elements ($R_+, R_-, T_+, T_-$). Just as the su(2)
algebra is defined by the commutation relations of its elements eq (\ref{eq:angularmoment}), so the
super-algebra is defined by the commutation or anti-commutation relations amongst all pairs of
its elements. First, there are the diagonal Weyl element relations:
\begin{eqnarray}
\left[ S_+, S_- \right] & = & 2 S_z, 
\nonumber \\
\{ R_+, R_-\} & = & A + S_z
\nonumber \\
\{ T_+, T_- \} & = & A - S_z
\label{eq:diagonalweyl}
\end{eqnarray}
As usual square brackets denote commutators; curly brackets, anti-commutators.
Next there are the off-diagonal Weyl commutation relations 
\bea
& [S_+, R_+] = - T_+, \hspace{1.5in} [S_-, R_+ ] = 0, \nonumber \\
& [S_+, R_-] = 0, \hspace{1.5in}  [S_-, R_-] = T_-, \nonumber \\
& [ S_+, T_+ ] = 0, \hspace{1.55in} [ S_-, T_+ ] = - R_+, \nonumber \\
& [ S_+, T_-] = R_-, \hspace{1.4in} [ S_-, T_-] = 0,
\label{eq:weylcommute}
\eea
and the off-diagonal Weyl anti-commutation relations
\bea
& \{ R_+, T_+ \} = 0, \hspace{1.5in}
 \{ R_-, T_+ \} = S_+, \nonumber \\
& \{ R_+, T_- \} = S_+, \hspace{1.5in}
 \{ R_-, T_- \} = 0.
\label{eq:weylanticommute}
\eea
The Cartan elements $A$ and $S_z$ commute with each other; 
$[A, S_z] = 0$. The final set of defining relations are the commutators of the
Weyl and Cartan elements:
\bea
& [S_+, S_z] = - S_+, \hspace{1.5in} [S_+, A] = 0, \nonumber \\
& [ S_-, S_z ] = S_-, \hspace{1.5in} [ S_-, A ] = 0, \nonumber \\
& [ R_+, S_z ] = \frac{1}{2} R_+, \hspace{1.5in} [ R_+, A ] = - \frac{1}{2} R_+, \nonumber \\
& [ R_-, S_z ] = - \frac{1}{2} R_-, \hspace{1.5in} [ R_-, A ] = \frac{1}{2} R_-, \nonumber \\
& [ T_+, S_z ] = - \frac{1}{2} T_+, \hspace{1.5in} [ T_+, A ] = - \frac{1}{2} T_+, \nonumber \\
& [ T_-, S_z ] = \frac{1}{2} T_- \hspace{1.5in} [ T_-, A ] = \frac{1}{2} T_-.
\label{eq:weylcartan}
\eea

These relations serve to define the algebra.

Now let us describe the action of the algebra elements on the states
of a super-multiplet. $S_+$ and $S_-$ simply raise and lower the 
$z$-component of the spin in either multiplet:
\begin{eqnarray}
S_+ | s, m > & = & (s - m)^{1/2} (s + m + 1)^{1/2} |s, m+1 >, 
\nonumber \\
S_+ | s - 1/2, m > & = & (s - 1/2 - m)^{1/2} (s + 1/2 + m)^{1/2} \nonumber \\
& & |s - 1/2, m + 1>,
\nonumber \\
S_- | s, m > & = & (s - m + 1)^{1/2} (s + m)^{1/2} |s, m - 1>, \nonumber \\
S_- | s - 1/2, m > & = & (s + 1/2 - m)^{1/2} (s - 1/2 + m)^{1/2} \nonumber \\
& & |s - 1/2, m - 1>.
\label{eq:s}
\end{eqnarray}
$R_+$ and $R_-$ switch states between multiplets
\begin{eqnarray}
R_+ | s, m > & = & (s + m)^{1/2} |s - 1/2, m - 1/2 >, 
\nonumber \\
R_+ | s - 1/2, m > & = & 0, 
\nonumber \\
R_- |s, m > & = & 0,
\nonumber \\
R_- |s - 1/2, m > & = & (s + 1/2 + m)^{1/2} |s, m + 1/2 >.
\nonumber \\
\label{eq:r} 
\end{eqnarray}
Note that $R_+$ lowers the $z$-component of spin by half when it changes from spin $s$ to
spin $s - 1/2$. $T_+$ and $T_-$ also switch states between multiplets
\begin{eqnarray}
T_+ |s, m > & = & (s - m)^{1/2} |s - 1/2, m + 1/2 >, 
\nonumber \\
T_+ |s - 1/2, m > & = & 0, 
\nonumber \\
T_- |s, m > & = & 0,
\nonumber \\
T_- |s - 1/2, m > &  = & (s + 1/2 - m)^{1/2} |s, m - 1/2 >,
\nonumber \\
\label{eq:t}
\end{eqnarray}
but whereas $R_+$ lowers the $z$-component by half, $T_+$ raises it. Finally the states of the
super-multiplet are eigenstates of $A$ and $S_z$
\begin{eqnarray}
A |s, m> & = & s | s, m >, \nonumber \\
A| s - 1/2, m > & = & (s + 1/2) |s - 1/2, m >, \nonumber \\
S_z |s, m > & = & m |s, m >, \nonumber \\
S_z |s - 1/2, m > & = & m |s - 1/2, m >.
\label{eq:a}
\end{eqnarray}
Thus the $A$ value distinguishes the multiplets; the $S_z$ value specifies the state within the
multiplet. Eqs (\ref{eq:s}), (\ref{eq:r}), (\ref{eq:t}) and (\ref{eq:a}) fully describe the action of the
super-algebra elements on the states of the super-multiplet. The normalization factors in these
equations follow inexorably from the commutation and anti-commutation relations that define the
super-algebra. Note that the action of $S_+, S_-$ and $S_z$ is exactly as one would expect
from the textbook theory of angular momentum; this is because these operators constitute an
su(2) subalgebra of our super-algebra. 


%
We can now write the $t-J$ Hamiltonian in supersymmetric form
\bea
H_{t-J} =  - \tau \sum_{{\mathbf r}, {\boldsymbol \delta}} 
[R_+ ({\mathbf r} + {\boldsymbol \delta}) R_- ({\mathbf r}) +
R_+ ({\mathbf r})  R_- ({\mathbf r} + {\boldsymbol \delta}) \nonumber \\
+ 
T_+ ({\mathbf r} + {\boldsymbol \delta}) T_- ({\mathbf r}) +
T_+ ({\mathbf r})  T_- ({\mathbf r} + {\boldsymbol \delta}) ]
\nonumber \\
+ J \sum_{{\mathbf r}, {\boldsymbol \delta}}
[ S_z ({\mathbf r}) S_z ({\mathbf r} + {\boldsymbol \delta}) - \{
A ({\mathbf r} ) - 2 s \} \{ A({\mathbf r} + {\boldsymbol \delta}) - 2 s \}
\nonumber \\
+ \frac{1}{2} S_+ ({\mathbf r} + {\boldsymbol \delta}) S_- ({\mathbf r}) +
\frac{1}{2} S_+ ({\mathbf r}) S_- ({\mathbf r} + {\boldsymbol \delta}) ].
\nonumber \\
\label{eq:tjsusy}
\eea
(For the traditional/non-supersymmetric expression, see for example section 3.2 of \cite{bib2}.)  
We assume the super-spins occupy the sites of a square lattice in a plane. 
The lattice position vectors are ${\mathbf r} = m a \hat{{\mathbf e}}_x + n a \hat{{\mathbf e}}_y$ 
where $m$ and $n$ are integers and the sum over ${\mathbf r}$ in eq (\ref{eq:tjsusy}) 
is over $m$ and $n$. ${\boldsymbol \delta}$ denotes the four nearest neighbor displacements
$\pm a \hat{{\mathbf e}}_x$ and $\pm a \hat{{\mathbf e}}_y$; the sum over ${\boldsymbol \delta}$
in eq (\ref{eq:tjsusy}) is over these four values. The super-spin operators at different sites are assumed
to commute and at a given site they are assumed to obey the super-algebra defined by
eqs (\ref{eq:diagonalweyl}), (\ref{eq:weylcommute}), (\ref{eq:weylanticommute}) and (\ref{eq:weylcartan}).
Thus the $t-J$ Hamiltonian couples super-spins
at neighboring sites. 
\par Finally a word about the symmetry of the Hamiltonian, $H_{t-J}$. The Heisenberg Hamiltonian
$H_F$ eq (\ref{eq:heisenbergferro}) has rotational symmetry. Formally this is demonstrated by
defining the total spin operators 
\begin{equation}
S_+^{{\rm tot}} = \sum_{{\mathbf r}} S_+ ({\mathbf r}) 
\label{eq:stot}
\end{equation}
(and $S_-^{{\rm tot}}$ and $S_z^{{\rm tot}}$ similarly) and verifying that 
$[H_F, S_+^{{\rm tot}} ] = 0$ (as well as 
$[H_F, S_-^{{\rm tot}} ] = 0$ and $[H_F, S_z^{{\rm tot}} ] = 0$).
In the same way we can define the total super-spin operator
\begin{equation}
R_+^{{\rm tot}} = \sum_{{\mathbf r}} R_+ ({\mathbf r}),
\label{eq:rtot}
\end{equation}
and similarly for all other elements of the super-algebra. For the $t-J$ Hamiltonian to be 
supersymmetric it would have to satisfy $[ H_{t-J}, R_+^{{\rm tot}} ] = 0$, $[ H_{t-J}, S_+^{{\rm tot}} ] = 0$
and so on for all eight elements of the super-algebra. This condition is not met except for special values
of the parameters $t$ and $J$, namely $|2\tau|=|J|$. The $t-J$ Hamiltonian is
certainly not supersymmetric for the experimentally relevant values. Thus although the Hamiltonian is
built out of supersymmetric algebra elements it is not generally supersymmetric. In this respect it is
similar to SUSY extensions of the standard model for which also super-symmetry is broken. 
%
\subsection{Dysonization of the $t-J$ Hamiltonian}
\label{sec-tjdysonized} 
Dyson's key insight was to define magnons as bosonic with respect to a non-standard inner product. For the $t-J$ model we wish to take that scheme one step further and define a `Dyson fermion' in addition to the Dyson bosons we have already alluded to. 
\par In order to represent the $t-J$ Hamiltonian in terms of Dyson bosons and fermions first let us consider
a single super-multiplet corresponding to the states at a single site. The basis states for a super-multiplet
that we have so far adopted are the $4s+1$ states $|s, m>$ and $|s-1/2, \mu>$ where $m = - s, \ldots, +s$
and $\mu = - (s - 1/2), \ldots, s - 1/2$.

\par Following Dyson we now introduce the alternative basis states
\begin{equation}
| u, 0 \rangle = F_{u,0} |s, - s + u \rangle, \; \; | u, 1 \rangle = F_{u,1} | s - 1/2, - (s - 1/2) + u \rangle
\label{eq:superdysonbasis}
\end{equation}
where $u = 0, \ldots, 2 s$ for the $|u, 0 \rangle$ states and 
$ u = 0, \ldots, 2s - 1$ for the $|u, 1 \rangle$ states. Thus $| 0, 0 \rangle$ corresponds to having a spin $s$
at the site that is maximally down; $|u, 0 \rangle$ corresponds to raising the spin $u$ times. Similarly
$|0, 1 \rangle$ corresponds to having a spin $s - 1/2$ at the site that is maximally down; $ | u, 1 \rangle$
corresponds to raising that spin $u$ times.  The states $|u, 0 \rangle$ are neutral; the states $| u, 1 \rangle$ correspond to having a net charge $+e$ on the site. Usually these sites are described as holons; in light of the supersymmetry discussion above, it seems natural to associate the charge with the presence of a non-Hermitian `Dyson fermion'.  Thus the filling fraction of Dyson fermions ({\em i.e.} the number of Dyson fermions per lattice site) is equal to the doping parameter $x$. 

The states in this basis are orthogonal to each other but not normalized:
\begin{equation}
\langle u, a | v, b \rangle = F_{u,a}^2 \delta_{ab} \delta_{uv}.
\label{eq:superorthogonal}
\end{equation}
The normalization factors $F_{u,a}$ are chosen judiciously:  
\begin{equation}
S^+ | u, a \rangle = \sqrt{2s} \sqrt{u + 1} | u + 1, a \rangle;
\label{eq:spluscondition}
\end{equation}
so as to maintain the action of $S^+$ as a bosonic raising operator. This is accomplished by 
defining
\begin{equation}
| u, a \rangle = \frac{1}{\sqrt{2s}^u} \frac{1}{\sqrt{u!}} (S^+)^u | 0, a \rangle,
\label{eq:uadefnd}
\end{equation}
which corresponds to the choice
\begin{equation}
F_{u,a} = 
\left( 1 - \frac{1}{2s} \right)^{1/2} 
\left( 1 - \frac{2}{2s} \right)^{1/2} \ldots
\left( 1 - \frac{ u - 1 + a}{2 s} \right)^{1/2}.
\label{eq:fua}
\end{equation}
The $|u, a\rangle$ basis is fully specified by eqs (\ref{eq:superdysonbasis}) and (\ref{eq:fua}) or equivalently
by eq (\ref{eq:uadefnd}). 

We may now determine the action of all the super-spin operators in this basis. The results are 
\begin{eqnarray}
S^+ | u, a \rangle & = & \sqrt{2s} \sqrt{u+1} | u + 1, a \rangle \nonumber \\
S^- | u, a \rangle & = & \sqrt{2s} \left[ 1 - \frac{u - 1 + a}{2 s} \right] u^{1/2} |u-1, a \rangle \nonumber \\
S_z |u, a \rangle & = & \left( - s + u + \frac{a}{2} \right) | u, a \rangle \nonumber \\
A |u, a \rangle & = & a | u, a \rangle, 
\label{eq:superbose}
\end{eqnarray}
for the commuting elements of the super-algebra, and 
\begin{eqnarray}
T^+ | u, 0 \rangle & = &  \sqrt{2 s} | u, 1 \rangle \nonumber \\
T^+ | u, 1 \rangle & = & 0 \nonumber \\
T^- |u, 0 \rangle & = & 0 \nonumber \\
T^- | u, 1 \rangle & = & \left[ 1 - \frac{u}{2s} \right] \sqrt{2s} |u, 0 \rangle \nonumber \\
R^+ |u, 0 \rangle & = & u^{1/2} | u - 1, 1 \rangle \nonumber \\
R^+ |u, 1 \rangle & = & 0 \nonumber \\
R^- |u, 0 \rangle & = & 0 \nonumber \\
R^- |u, 1 \rangle & = & (u + 1)^{1/2} |u + 1, 0 \rangle, 
\label{eq:superfermi}
\end{eqnarray}
for the anti-commuting elements. 
Now consider a different Hilbert space inhabited by a single Bose creation and annihilation operator
pair $(b, b^\dagger)$ and a Fermi pair $(a, a^\dagger)$ that satisfy the canonical commutation relations
\begin{eqnarray}
[ b, b^\dagger ] & = & 1, \nonumber \\
\{ a, a^\dagger \} & = & 1, \; a^2 = a^{\dagger 2} = 0.
\label{eq:canonicalbosefermi}
\end{eqnarray}
We also suppose $[ a, b ]  =  [ a^\dagger, b ] = [ a, b^\dagger ] = [ a^\dagger, b^\dagger ] = 0$.
The creation and annihilation operators are adjoints of each other under the kinematical inner product
in this Hilbert space. One can show inexorably from these assumptions that the basic states of this 
Hilbert space are $|u, 0 )$ and $|u, 1)$ where $u = 0, 1, 2, \ldots$ The state $|0, 0)$ has the defining
characteristic 
\begin{equation}
b | 0, 0 ) = a | 0, 0 ) = 0;
\label{eq:bosefermivacuum}
\end{equation}
it contains neither a $b$ boson not an $a$ fermion. The state
\begin{equation}
|u, 0 ) = \frac{1}{\sqrt{u!}} (b^\dagger)^u | 0, 0 )
\label{eq:u0bose}
\end{equation}
contains $u$ bosons and no fermions. The state
\begin{equation}
| u, 1 ) = \frac{1}{\sqrt{u!}} (b^\dagger)^u a^\dagger |0, 0) 
\label{eq:u1fermi}
\end{equation}
contains $u$ bosons and one fermion. These states are orthonormal under the kinematic inner product
\begin{equation}
(u, a | v, b )_{{\rm kin}} = \delta_{u,v} \delta_{a,b}. 
\label{eq:superkininner}
\end{equation}

We now establish the following mapping between the states of a supermultiplet and the bose-fermi
Hilbert space discussed above. The mapping is
\begin{equation}
| u, a \rangle \rightarrow |u, a) 
\label{eq:supermap}
\end{equation}
Here $u = 0, \ldots, 2s$ for $a = 0$ and $u = 0, \ldots, 2s - 1$ for $a = 1$. States with more bosons have no
counter-part in the super-spin space. 

As before this correspondence exports a dynamical inner product to the Bose-Fermi Hilbert space
\begin{equation}
(u, a| v, b)_{{\rm dyn}} = F_{ua}^2 \delta_{uv} \delta_{ab}. 
\label{eq:superdynamical}
\end{equation}
We assume $F_{u,0} = 0$ for $u > 2 s$ and $ F_{u, 1} = 0$ for $ u > 2 s -1$. Thus states with a greater
number of bosons are weightless. 

The mapping eq (\ref{eq:supermap}) also allows us to establish a correspondence between 
super-spin and bose and fermi operators. The correspondence follows from eqs (\ref{eq:superbose})
and eq (\ref{eq:superfermi}) and is as follows 

\bea
S^+ \rightarrow \sqrt{2s} b^\dagger, \hspace{1.5in} S^- \rightarrow \left[ 1 - \frac{b^\dagger b + a^\dagger a}{2 s} \right] 
\sqrt{2 s} b \nonumber \\
S_z \rightarrow ( - s + b^\dagger b + \frac{1}{2} a^\dagger a ), \hspace{1.5in} A \rightarrow a^\dagger a 
\nonumber \\
T^+ \rightarrow \sqrt{2s} a^\dagger   \hspace{1.5in} T^- \rightarrow \left[ 1 - \frac{b^\dagger b + a^\dagger a}{2 s} \right]
\sqrt{2s} a \nonumber \\
R^+ \rightarrow b a^\dagger \hspace{1.5in} R^- \rightarrow a b^\dagger.
\label{eq:firstcorrespondence}
\eea

\subsubsection {Ferromagnetic t-J model}
Now let us consider the $t-J$ model eq (3-37). For the cuprates
we are interested in anti-ferromagnetic coupling $(J > 0)$ but it is instructive to first 
consider the case of ferromagnetic coupling, $J < 0$.

We introduce a single boson $ b({\mathbf r}), b^\dagger ({\mathbf r})$ and a single fermion
$ a({\mathbf r}), a^\dagger ({\mathbf r}) $ at each site of the lattice. Bose and Fermi creation and annihilation
operators at the same site are taken to be adjoints of each other under the kinematic inner product.
Using the correspondence between super-spin operators and bose and fermi operators, eq 
(\ref{eq:firstcorrespondence}), we may write the $t-J$ Hamiltonian as
\begin{eqnarray}
H_{t-J} & = & - 2 \tau s \sum_{{\mathbf r}, {\boldsymbol \delta}} 
[ a^\dagger ({\mathbf r} + {\boldsymbol \delta} ) a ( {\mathbf r} ) + a^\dagger ({\mathbf r}) a ({\mathbf r} + {\boldsymbol 
\delta}) ]
\nonumber \\
& & + \frac{1}{2} J s \sum_{{\mathbf r}, {\boldsymbol \delta}} a^\dagger ({\mathbf r}) a({\mathbf r})
\nonumber \\
& & + \frac{1}{2} J s \sum_{{\mathbf r}, {\boldsymbol \delta} } [ \ b^\dagger ({\mathbf r}) b ({\mathbf r}) - 
b^\dagger ( {\mathbf r} + {\boldsymbol \delta}) b ({\mathbf r}) ] + \ldots
\label{eq:leadingtjferro}
\end{eqnarray}
In eq (\ref{eq:leadingtjferro}) we have written out the leading quadratic term in the Dyson representation
of the ferromagnetic $t-J$ Hamiltonian. At this level, it is a theory of non-interacting bosonic spin-waves
(``magnons'') and fermions with charge $+e$ (``magninos''). 

The interaction terms that were omitted in eq (\ref{eq:leadingtjferro}) and are presumably small in this representation, 
are given by
\begin{eqnarray}
H_{{\rm int}} & = &  - \tau 
\sum_{{\mathbf r}, {\boldsymbol \delta}} b^\dagger ({\mathbf r}) b ({\mathbf r}) a({\mathbf r}) a^\dagger ({\mathbf r}
+ {\boldsymbol \delta} ) \nonumber \\
& & - \tau \sum_{{\mathbf r}, {\boldsymbol \delta} } 
[ b ({\mathbf r} + {\boldsymbol \delta} ) b^\dagger ( {\mathbf r} ) a^\dagger ( {\mathbf r} + {\boldsymbol \delta} )
a ( {\mathbf r} ) + 
b^\dagger ({\mathbf r}) b( {\mathbf r} + {\boldsymbol \delta})
a^\dagger ({\mathbf r}) a ({\mathbf r} + {\boldsymbol \delta}) ]
\nonumber \\
& & - \frac{J}{2} \sum_{{\mathbf r}, {\boldsymbol \delta}} [ b^\dagger ({\mathbf r}) b( {\mathbf r} ) + \frac{1}{2}
a^\dagger ({\mathbf r}) a({\mathbf r}) ] [ b^\dagger ({\mathbf r} + {\boldsymbol \delta}) b ( {\mathbf r} + {\boldsymbol 
\delta} ) + \frac{1}{2} a^\dagger ( {\mathbf r} + {\boldsymbol \delta} ) a ({\mathbf r} + {\boldsymbol \delta} ) ]
\nonumber \\
& & + \frac{J}{4} \sum_{{\mathbf r}, {\boldsymbol \delta} } 
[ a^\dagger ({\mathbf r}) a ({\mathbf r}) + b^\dagger ( {\mathbf r} ) b ({\mathbf r}) ]
[ b^\dagger ( {\mathbf r} + {\boldsymbol \delta} ) b ({\mathbf r}) ]. \nonumber \\
\label{eq:hint}
\end{eqnarray}
The full $t-J$ Hamiltonian, $H_{t-J}$ is not self-adjoint under the kinematic inner product $(H_{t-J}^\dagger \neq
H_{t-J}$); however it is self-adjoint under the dynamical inner product, $H_{t-J}^\star = H_{t-J}$.
\subsubsection{Anti-ferromagnetic t-J model} 
For the anti-ferromagnetic $t-J$ model, as for the Heisenberg anti-ferromagnet,
it is convenient to imagine a pair of interpenetrating square lattices. The $t-J$ Hamiltonian may then be
re-written 
\begin{eqnarray}
H_{t-J} & = & - \tau 
\sum_{{\mathbf r}, {\boldsymbol \delta} } [ R_+^{(2)} ( {\mathbf r} + {\boldsymbol \delta}) R_-^{(1)} ({\mathbf r}) 
+
R_+^{(1)} ({\mathbf r}) R_-^{(2)} ({\mathbf r} + {\boldsymbol \delta} ) ]
\nonumber \\ 
& & - \tau
\sum_{{\mathbf r}, {\boldsymbol \delta} }
[
T_+^{(2)} ({\mathbf r} + {\boldsymbol \delta} ) T_-^{(1)} ({\mathbf r}) +
T_+^{(1)} ({\mathbf r} ) T_-^{(2)} ({\mathbf r} + {\boldsymbol \delta} ) ]
\nonumber \\
& & + J 
\sum_{ {\mathbf r}, {\boldsymbol \delta} } [ 
S_z^{(1)} ({\mathbf r}) S_z^{(2)} ({\mathbf r} + {\boldsymbol \delta} ) - 
\{ A^{(1)} ({\mathbf r}) - 2 s \} \{ A^{(2)} ({\mathbf r} + {\boldsymbol \delta} ) -  2 s \} ]
\nonumber \\
& & + \frac{J}{2} \sum_{ {\mathbf r}, {\boldsymbol \delta} } 
[ S_+^{(2)} ({\mathbf r} + {\boldsymbol \delta} ) S_-^{(1)} ({\mathbf r}) + S_+^{(1)} ({\mathbf r})
S_-^{(2)} ({\mathbf r} + {\boldsymbol \delta}) ].
\label{eq:aftj}
\end{eqnarray}
The sum over ${\mathbf r}$ in eq (\ref{eq:aftj}) extends over the sites of the first lattice; 
the sum over ${\boldsymbol \delta}$ extends over the four nearest neighbors of each site.
The superscripts $^{(1)}$ and $^{(2)}$ over the super-spin operators serve to remind us
that the spin is on lattice one or lattice two respectively. 

At least for light doping it makes sense to assume that the N\'{e}el
 state is a good starting point for the
ground state of the $t-J$ model. In the N\'{e}el
 state the spin is maximally down at each site of the
first lattice; it is maximally up at each site of the second lattice. The magnitude of the spin is
$s-1/2$ at sites occupied by a hole. It is $s$ at all other sites. In representing the N\'{e}el
 state
in terms of Dyson bosons and fermions we shall take the N\'{e}el
 state to have zero bosons
and to have Dyson fermions at all the sites with holes. 

To this end we establish a second mapping between the states of a single super-spin and 
the Hilbert space of a single boson and fermion. In this mapping we identify the states with
spin maximally up as the state with zero bosons whereas before we had assigned this
part to spin maximally down. Thus we introduce the basis
\begin{eqnarray}
| u, 0 \rangle & = & G_{u, 0} | s, s - u \rangle \nonumber \\
| u, 1 \rangle & = & G_{u,1} | s - 1/2, s-1/2 - u \rangle
\label{eq:secondsuperbasis}
\end{eqnarray}
in place of eq (\ref{eq:superdysonbasis}).
This time we choose 
\begin{equation}
G_{u,a} = \left( 1 - \frac{1}{2s} \right)^{-1/2} \left( 1 - \frac{2}{2s} \right)^{-1/2} \ldots
\left( 1 - \frac{ u - 1 + a }{2 s} \right)^{-1/2}.
\label{eq:secondfua}
\end{equation}
As before we then establish a mapping $|u, a \rangle$ between the states of the
super-spin and the states $|u, a)$ of a bose-fermi system. By virtue of this correspondence
we obtain a second mapping between super-spin and bose and fermi operators: 
\bea
& S^+ \rightarrow \sqrt{2s} b \hspace{1.5in} S^- \rightarrow \sqrt{2s} b^\dagger \left(1 - \frac{b^\dagger b + a^\dagger a}{2s} \right),
\nonumber \\
& S_z \rightarrow (s - b^\dagger b - \frac{1}{2} a^\dagger a ), \hspace{.6in} A \rightarrow a^\dagger a,
\nonumber \\
& R^+ \rightarrow \sqrt{2s} \left( 1 - \frac{b^\dagger b}{2 s} \right), \hspace{.6in} R^- \rightarrow \sqrt{2s} a,
\nonumber \\
& T^+ \rightarrow b a^\dagger, \hspace{1.5in} T^- \rightarrow b^\dagger a.
\label{eq:secondsupercorrespondence}
\eea

In  order to write the $t-J$ Hamiltonian in terms of bosons and fermions we use the first correspondence
eq (\ref{eq:firstcorrespondence}) on the first lattice and the second correspondence eq (\ref{eq:secondsupercorrespondence})
on the second
lattice. Keeping the leading terms up to cubic order we obtain a novel representation of the $t-J$ 
Hamiltonian in terms of bosons and fermions:
\begin{eqnarray}
h_{t-J} & = & J s 
\sum_{{\mathbf r}, {\boldsymbol \delta}}
[ b_1^\dagger ( {\mathbf r} ) b_1 ({\mathbf r}) + 
b_2^\dagger ({\mathbf r} +{\boldsymbol \delta} ) b_2 ({\mathbf r} + {\boldsymbol \delta}) ] 
\nonumber \\
& & + J s 
\sum_{ {\mathbf r}, {\boldsymbol \delta} }
[ a_1^\dagger ({\mathbf r}) a_1 ({\mathbf r}) +
a_2^\dagger ({\mathbf r} + {\boldsymbol \delta}) a_2 ({\mathbf r} + {\boldsymbol \delta} )]
\nonumber \\
& & + J s \sum_{{\mathbf r}, {\boldsymbol \delta}}
[ b_1 ({\mathbf r}) b_2 ({\mathbf r} + {\boldsymbol \delta} ) + b_2^\dagger ({\mathbf r} + {\boldsymbol \delta})
b_1^\dagger ({\mathbf r} ) ]
\nonumber \\
& & - \tau \sqrt{2s} 
\sum_{{\mathbf r}, {\boldsymbol \delta}} [
a_1^\dagger ( {\mathbf r} ) a_2 ({\mathbf r} + {\boldsymbol \delta}) b_1 ({\mathbf r}) + 
a_2^\dagger ({\mathbf r} + {\boldsymbol \delta}) a_1 ({\mathbf r}) b_1^\dagger ({\mathbf r}) ]
\nonumber \\
& & - \tau \sqrt{2s} \sum_{{\mathbf r}, {\boldsymbol \delta}} 
[ a_1^\dagger ({\mathbf r}) a_2 ({\mathbf r} + {\boldsymbol \delta}) b_2^\dagger({\mathbf r} + {\boldsymbol \delta}) +
a_2^\dagger ({\mathbf r} + {\boldsymbol \delta}) a_1 ({\mathbf r}) b_2 ({\mathbf r} + {\boldsymbol \delta})] + \ldots
\nonumber \\
\label{eq:tjcubic}
\end{eqnarray}
The remaining interaction terms which are quartic and quintic are presumably small in this representation, but we leave these calculations for future work, as our purpose here is simply to construct the relevant formalism. 

The essential physics of the $t-J$ model in this regime is thus revealed to be that of charged non-Hermitian fermions hopping in a background of spin-waves. This represents a novel formulation of the problem of weakly doped anti-ferromagnets that has been extensively studied beginning with the seminal work of Kane {\em et al} \cite{bib8}. It is a tantalizing possibility that the non-Hermitian quasi-particles defined here may shed new light on the underlying physics of high $T_c$ materials. 
 
\section{Conclusion}
The applications of non-Hermitian quantum mechanics may extend beyond the realm of fundamental physics into the emergent world of condensed matter. Dyson unwittingly discovered non-Hermitian quantum mechanics in 1956 \cite{bib5} when he found that high precision calculations of interacting spin waves in a ferromagnet were facilitated by use of a non-Hermitian Hamiltonian. Dyson's 
technique of defining quasi-particles with respect to a non-standard inner product allows for a novel way of writing the t-J Hamiltonian; this new form of the t-J Hamiltonian may prove more wieldy to calculations and even shed some light on the physics that underlies high temperature superconductivity, arguably the most outstanding problem in theoretical condensed matter physics \cite{bib9}.


\begin{thebibliography}{99}
\bibitem{bib1}  Anderson, P.W. 1997 {\em Basic Notions of Condensed Matter Physics} Westview Press, 
2nd edition. 
\bibitem{bib2} Auerbach, A. 1994 \textit{Interacting Electrons and Quantum Magnetism}, Springer, New York. 
\bibitem{bib2a} Bares,P.A.  \& Blatter, G. 1990 Supersymmetric t-J model in one dimension - separation of spin and charge, \textit{Phys. Rev. Lett.} {\bf 64}, 2567. 1990. 
\bibitem{bib3} Bender, C.M.  2007 Making Sense of Non-Hermitian Hamiltonians,
\textit{Rept. Prog. Phys.} {\bf 70}, 947. 
\bibitem{bib4} Bender, C.M. \&  Boettcher, S. 1998  Real Spectra in 
Non-Hermitian Hamiltonians Having PT Symmetry, \textit{Phys Rev Lett} {\bf 80}, 
5243. 
\bibitem{bib5} Dyson, F.J. 1956  General Theory of Spin-Wave Interactions, \textit{Phys
Rev} {\bf 102}, 1217. 
\bibitem{bib6}  Harris, A.B., Kumar, D. , Halperin, B.I.  \& Hohenberg,P.C.   1971 Dynamics of an Antiferromagnet at Low Temperatures: Spin-Wave Damping and Hydrodynamics,
\textit{Phys. Rev. B3}, 961. 
\bibitem{bib7} Jones-Smith, K. 2010 Non-Hermitian Quantum Mechanics, \textit{Ph.D Thesis}, Case Western Reserve University. 
\bibitem{bib8} Kane,C.L., Lee, P.A.  \& Read,N. 1989  Motion of  a single hole in a quantum antiferromagnet, \textit{Phys. Rev B} {\bf 39}, 6880.
\bibitem{bib9} Leggett,A.J. 2006  What DO we know about high $T_c$?,  \textit{Nature Physics}, {\bf 2}, 134. 
\bibitem{bib10} Manousakis,E.  1991 The spin 1/2 Heisenberg antiferromagnet on a square lattice and its application to the cuprous oxides. \textit{Rev. Mod. Phys.} 63, 1. 
\bibitem{bib11} Weigmann, P.C. 1988 Superconductivity  in strongly correlated electronic systems and confinement vs de-confinement phenomena, \textit{Phys. Rev. Lett.} 60, 821.  

\end{thebibliography}
\end{document}